\documentclass[conference]{IEEEtran}

\usepackage{cite}
\usepackage[numbers,sort&compress]{natbib}

\usepackage[utf8]{inputenc}
\usepackage{amsmath,amssymb}
\usepackage{graphicx}
\usepackage{bm}
\usepackage{enumerate}
\usepackage{comment}
\usepackage{xcolor}
\usepackage{url}
\usepackage{color,soul}
\usepackage{subcaption}
\usepackage{float}

\definecolor{light-gray}{gray}{0.8}

\usepackage{amsmath,amssymb,amsfonts}
\usepackage{algorithmic}
\usepackage{graphicx}
\usepackage{textcomp}
\usepackage{xcolor}
\usepackage{subcaption}

\def\BibTeX{{\rm B\kern-.05em{\sc i\kern-.025em b}\kern-.08em
    T\kern-.1667em\lower.7ex\hbox{E}\kern-.125emX}}

\makeatletter
\newcommand{\linebreakand}{%
  \end{@IEEEauthorhalign}
  \hfill\mbox{}\par
  \mbox{}\hfill\begin{@IEEEauthorhalign}
}

\begin{document}

\title{A Comparative Study on Enhancing Prediction in Social Network Advertisement through Data Augmentation\\}

\author{
\small % Set font size to 10pt

\begin{tabular}[t]{c@{\extracolsep{8em}}c} 

1\textsuperscript{st} Qikai Yang & 2\textsuperscript{nd} Panfeng Li \\
\textit{Department of Computer Science} & \textit{Department of Electrical and Computer Engineering} \\
\textit{University of Illinois at Urbana-Champaign} & \textit{University of Michigan} \\
Urbana, USA & Ann Arbor, USA \\
qikaiy2@illinois.edu & pfli@umich.edu\\

\\

3\textsuperscript{rd} Xinhe Xu & 4\textsuperscript{th} Zhicheng Ding  \\
\textit{Department of Computer Science} & \textit{Fu Foundation School of Engineering and Applied Science} \\
\textit{University of Illinois at Urbana-Champaign} & \textit{Columbia University} \\
Urbana, USA & New York, USA \\
xinhexu2@illinois.edu & zhicheng.ding@columbia.edu \\

\\
5\textsuperscript{th} Wenjing Zhou & 6\textsuperscript{th} Yi Nian \\
\textit{Department of Statistics} & \textit{Department of Computer Science} \\
\textit{University of Michigan} & \textit{University of Chicago} \\
Ann Arbor, USA & Chicago, USA \\
wenjzh@umich.edu & nian@uchicago.edu \\
\end{tabular}

}

\maketitle

\begin{abstract}
In the ever-evolving landscape of social network advertising, the volume and accuracy of data play a critical role in the performance of predictive models. However, the development of robust predictive algorithms is often hampered by the limited size and potential bias present in real-world datasets. This study presents and explores a generative augmentation framework of social network advertising data. Our framework explores three generative models for data augmentation - Generative Adversarial Networks (GANs), Variational Autoencoders (VAEs), and Gaussian Mixture Models (GMMs) - to enrich data availability and diversity in the context of social network advertising analytics effectiveness. By performing synthetic extensions of the feature space, we find that through data augmentation, the performance of various classifiers has been quantitatively improved. Furthermore, we compare the relative performance gains brought by each data augmentation technique, providing insights for practitioners to select appropriate techniques to enhance model performance. This paper contributes to the literature by showing that synthetic data augmentation alleviates the limitations imposed by small or imbalanced datasets in the field of social network advertising. At the same time, this article also provides a comparative perspective on the practicality of different data augmentation methods, thereby guiding practitioners to choose appropriate techniques to enhance model performance.
\end{abstract}


\section{Introduction}
In the age of digital marketing, social network advertising has become a key component of strategic communication and audience engagement. Various studies have been conducted on social network analysis ~\cite{1}. However, in this field, the effectiveness of predictive analytics relies on the availability of large and diverse datasets ~\cite{2}. This phenomenon is also commonly seen in many other scientific fields and many different studies ~\cite{3,4,5,6}. However, the complexity and time sensitiveness of social network data is different from other domains, leading to challenging problems. In many cases, the available datasets are either limited in size or biased, which can cause regression in the predictive models. Although there are some solutions to reduce the data bias in machine learning ~\cite{25}, this study aims to address these challenges by exploring data augmentation as a strategy to enhance and enlarge the data, thereby improving predictive performance.

In addition to the inherent difficulties posed by social data, our study also tries to address the importance of this problem by exploring the transformative potential of machine learning models in driving advertising effectiveness. Accurate prediction of user responses to ads can lead to success in personalized marketing, optimized ad delivery, and better ad budget allocation \cite{6}. However, the performance of these models is often limited by the data on which they are trained. Robust datasets are the cornerstone of any effective machine learning application, yet the scarcity of such data is a common obstacle to overcome ~\cite{1,7}.

The underlying hypothesis of this study is that synthetic data augmentation can serve as an effective solution to overcome the limitations of datasets. By generating new data points that mimic real data, the model can potentially learn more generalizable patterns, thereby improving accuracy when making predictions on unseen data. Similar methods have been taken in other fields such as CT segmentations ~\cite{5,8}, natural language processing \cite{9}, real-time systems \cite{15,16}, etc. Hence, our study endeavors to translate successful methodologies from other domains to the realm of social network advertising.

To testify our hypothesis, we employ three advanced data augmentation techniques: Generative Adversarial Networks (GANs) \cite{13,10}, Variational Autoencoders (VAEs) \cite{14,11}, and Gaussian Mixture Models (GMMs). Each technique has a unique approach to generating synthetic data, and its effectiveness in the field of social network advertising has not been extensively compared in previous research.

Our research revolves around two main questions:

How does synthetic data using GAN, VAE, and GMM affect the performance of various classifiers in the field of social network advertising.

Out of these data augmentation techniques, which one provides the best improvement in forecast accuracy.

This study systematically explores these topics. By systematically augmenting a real social network advertising dataset and evaluating the impact on model performance, we reveal the potential of data augmentation to revolutionize the field of advertising analytics. This introduction sets the stage for a comprehensive study that not only covers the technical areas of synthetic data generation, but also critically explores its practical implications within the broader context of social network advertising.

\section{Methods}
This section demonstrates the thorough methodology employed in this study to investigate how data augmentation affects the predictive accuracy of machine learning models. It encompasses the data preparation phase, the techniques utilized for generating synthetic data, and the detailed procedures for model training and evaluation.

\subsection{Data Collection and Preprocessing}
The primary data contains various characteristics that reflect user behavior on social networks and responses to advertisements. Initial preprocessing includes data cleaning, normalization, and encoding of categorical variables to prepare the data for the augmentation and training stages. Then, the dataset is split into training and test sets, maintaining the distribution to reflect the hierarchical structure of the original data. The original data includes 400 users.

\subsection{Framework}
\begin{figure}[t]
    \centering{
    \includegraphics[width=\columnwidth]{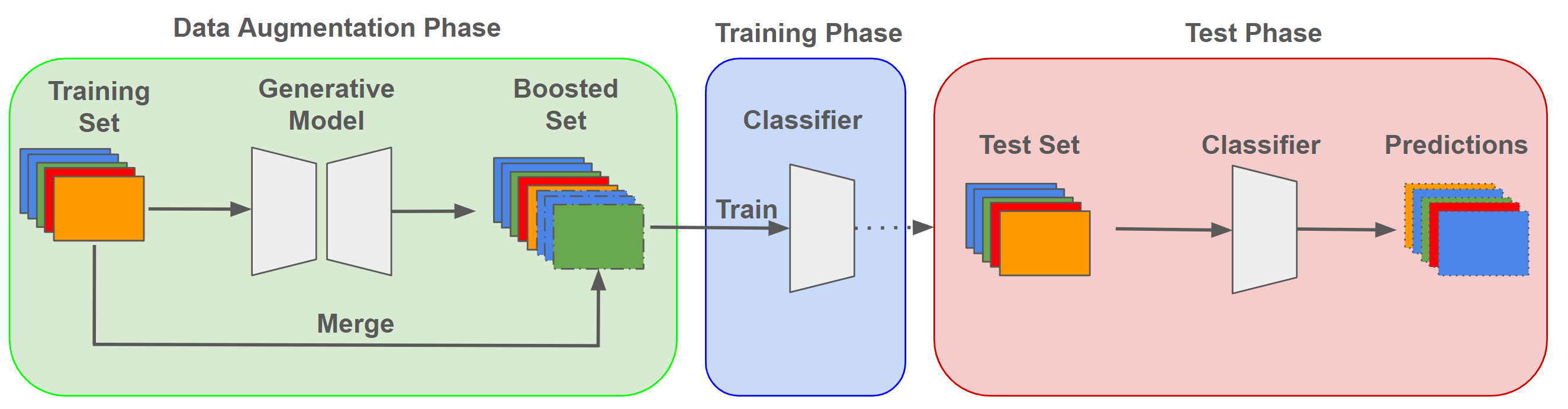}
    \caption{Data Augmentation Framework. In addition to the regular machine learning train phase and test phase, we add a data augmentation phase at the very beginning to expand the training set by utilizing generative models.} 
    \label{Architecture}}
\end{figure}

To fully test our generative data augmentation techniques, our framework contains three phases - data augmentation phase, training phase, and test phase, as indicated in Fig. \ref{Architecture}. 

\subsection{Synthetic Data Generation}
\begin{figure}[t]
    \centering
    \includegraphics[width=\columnwidth]{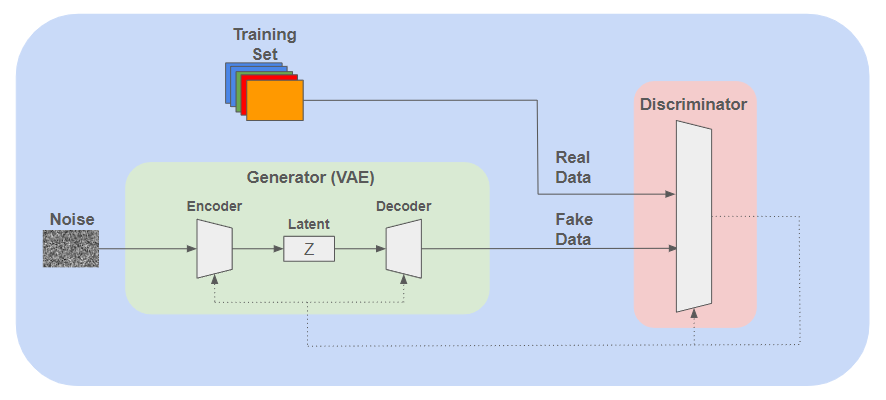}
    \caption{GAN with VAE as generator. We embed VAE to our GAN model as generator and regular multi-layer linear network as discriminator.} 
    \label{fig:vaegan}
\end{figure}

Synthetic data augmentation is performed using three generative models: GAN, VAE, and GMM. Applying three models respectively generated additional 200 users as new data instances:

GAN: Within a GAN framework, two neural networks engage in a competitive zero-sum game, wherein the advancement of one network corresponds to the setback of the other. The objective of GAN training is to produce data that closely resembles the distribution of the original training data. Once the generator reaches a suitable stage, it generates synthetic data, which is then combined with the original training data to construct a unified dataset. In our investigation, we employ a VAE as the generator and a multi-layer linear network as the discriminator as indicated in Fig. \ref{fig:vaegan}.

VAE: A variational autoencoder consists of a generative model equipped with a prior and noise distribution, each serving distinct roles. Employing an encoder-decoder architecture, VAE is designed to acquire a latent representation of the data, enabling the generation of new instances through sampling and decoding. In our research, we opt for a multi-layer linear network to serve as both the encoder and decoder.

GMM: A Gaussian mixture model is a soft clustering technique used in unsupervised learning to determine the probability that a given data point belongs to a cluster. Our GMM is used to study the distribution of training data, assuming that the data points are generated by a Gaussian distribution with multiple unknown parameters. The boosted samples are sampled from estimated Gaussian  distributions.

\subsection{Classification Model Training}
Six classifiers are explored in this study: decision trees, dense neural networks, K-nearest neighbor algorithm (KNN), logistic regression, and two variants of support vector machines, SVM with linear kernel and SVM with radial basis function (RBF) kernel. These models are chosen due to their ability to capture the diversity of different patterns in the data:

Decision tree: We set the Gini index as a splitting criterion and optimize the maximum depth through cross-validation to prevent overfitting.

Dense neural network: We build a multi-layer linear network, use the ReLU activation function, and train using the Adam optimization method. Similar settings have been used in previous research \cite{13,17,18}.

KNN: We determine the optimal number of neighbors through cross-validation and adjust distance weighting to improve the 
classification on minority classes.

Logistic regression: We adjust regularization strength and solver type to handle nonlinear relationships in our data. This model is widely used in various fields.

SVM (linear SVC and SVC): For linear support vector classifiers (SVC), we adjust the regularization parameters; for SVC, we use the RBF kernel and optimize the regularization parameters through cross validation.

Each model is first trained on the original train set and then on the boosted training set. This allows comparison of performance metrics across models and datasets.

\subsection{Performance Evaluation}
The primary focus of our model performance evaluation lies in assessing classification accuracy, F1 score, and AUC score, as they stand as the foremost criterion for gauging a model's predictive capacity.

\subsection{Ethical Considerations}
When generating synthetic data, we ensure that bias is not introduced or exacerbated. The distribution of synthetic data is closely monitored and the ethical implications of using such data are critically assessed throughout the study.

By strictly following this methodological framework, this study ensures the reliability of its findings and provides a template for future data enhancement and machine learning research in the field of social network advertising.

\section{Experiments}
\begin{figure}[t]
    \centering
    \includegraphics[width=\columnwidth]{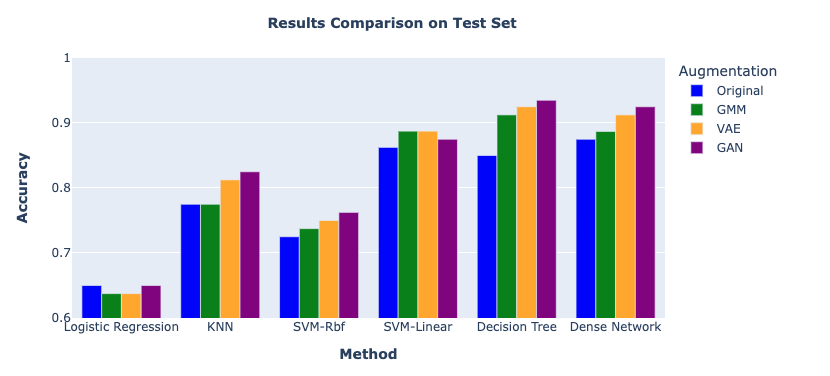}
    \caption{Classification accuracy on test set with the combination of all generative models and all classification models. } 
    \label{fig:result}
\end{figure}

Table \ref{tab:acc_f1} and Fig. \ref{fig:result} present the results obtained by applying different machine learning classifiers to the original and enhanced social network advertising datasets. The findings are based on a comparative analysis of different model performance metrics. 

\begin{table*}[h] % 'h' places the table here, within the text
    \centering % Centers the table
    
    \begin{tabular}{|c|c|c|c|c|c|c|c|c|c|c|c|c|} 
        \hline
        \multicolumn{1}{|c|}{} & \multicolumn{12}{|c|}{Classification Model} \\ \hline 
        \multicolumn{1}{|c|}{Boost Option} & \multicolumn{2}{c|}{Decision Tree} & \multicolumn{2}{c|}{KNN} & \multicolumn{2}{c|}{Logistic Regression} & \multicolumn{2}{c|}{SVM (RBF)} & \multicolumn{2}{c|}{SVM Linear} & \multicolumn{2}{c|}{Dense Network}\\ \hline
        Metrics & Acc & F1 & Acc & F1 & Acc & F1 & Acc & F1 & Acc & F1 & Acc & F1\\ \hline
        No boost & 0.85 & 0.79 & 0.78 & 0.63  & \textbf{0.64} & 0 & 0.73 & 0.45  & 0.86 & 0.78  & 0.88 & 0.80 \\ % Existing data
        GMM & 0.91 & 0.87 & 0.81 & 0.69  & 0.44 & 0.35 & 0.74 & 0.49  & \textbf{0.89} & \textbf{0.82}  & 0.91 & 0.87 \\ % Existing data
        VAE & 0.93 & 0.90 & \textbf{0.83} & \textbf{0.72}  & \textbf{0.64} & 0.06 & 0.75 & 0.52  & \textbf{0.89} & \textbf{0.82}  & \textbf{0.93} & \textbf{0.90} \\
        GAN & \textbf{0.94} & \textbf{0.92} & 0.78 & 0.65  & 0.48 & \textbf{0.36} & \textbf{0.76} & \textbf{0.60}  & 0.88 & 0.80  & 0.89 & 0.82 \\
        \hline
    \end{tabular}
    \caption{Classification Accuracy \& F1 Score}
    \label{tab:acc_f1}
\end{table*}

\subsection{Model Performance on Original Data}
The baseline performance of each classifier (decision tree, dense neural network, K-nearest neighbor algorithm (KNN), logistic regression, and linear SVM and SVM with RBF kernel) is established by training on the original dataset. The obtained performance metrics are used as a comparison baseline to evaluate the impact of data augmentation.

\subsection{Model Performance on Augmented Data}
\begin{figure}[t]
    \centering
    \includegraphics[width=\columnwidth]{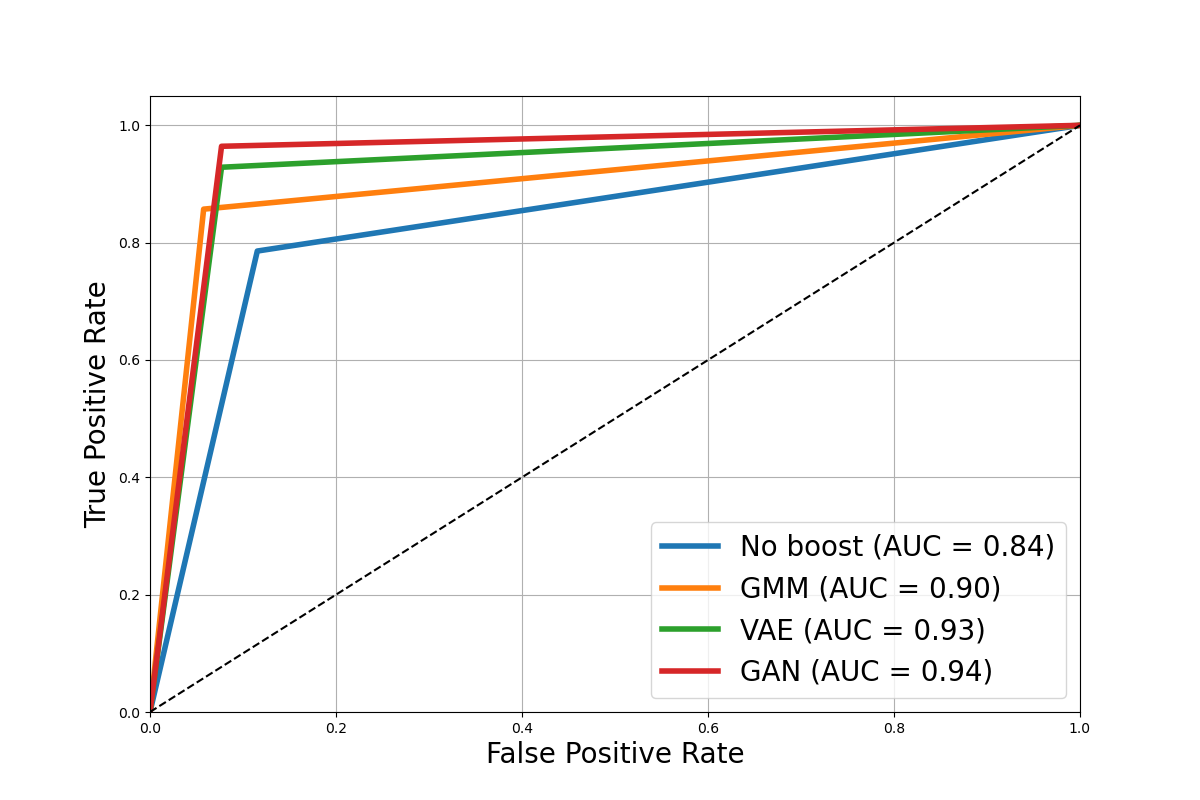}
    \caption{Result Comparison} 
    \label{fig:roc}
\end{figure}

Subsequently, the classifiers are trained on the augmented datasets generated by the three generative models. The performance metrics of these classifiers show several trends:

\underline{Decision Trees}: The three data augmentation methods all bring benefits to classification in decision trees. Among three methods, GAN outperforms the others. From Fig. \ref{fig:roc}, we can see that data augmentation from GAN successfully improves the AUC score from 0.84 to 0.94.

\underline{KNN}: The enhancement in KNN classifiers varies across different augmentation methods. While improvements in prediction accuracy are observed with GMM and VAE enhancements, the impact of GAN augmentation is negligible.

\underline{Logistic regression}: Logistic regression models trained on all augmented data show no improvements in accuracy but decent increase in F1 score.

\underline{SVM (linear SVC and SVM with RBF kernel)}: The accuracy of the SVM model on the augmented data is generally improved. All three data augmentation methods show similar and effective improvements.

\underline{Dense Neural Networks}: Dense neural network classifiers benefit greatly from augmented datasets, with VAE-generated data showing the best improvements across all metrics, which can be reflected from both accuracy and F1 score.

\subsection{Comparative Analysis}
The comparative analysis aims to determine the relative effectiveness of each data augmentation technique in improving the performance of various classifiers: 

\underline{GAN vs. VAE vs. GMM}: In general, VAE and GAN augmented data tend to result in more significant performance improvements across all models compared to GMM. Notably, within the realm of dense neural network classifiers, VAE consistently demonstrates superior performance.

\underline{Overfitting analysis}: An investigation into possible overfit- ting caused by synthetic data found that although the training accuracy of some models is slightly improved, this doesn’t not adversely affect the generalization ability of the test set.

\subsection{Interpretation of Results}
The improvement in model performance demonstrates that synthetic data augmentation can effectively cope with data constraints in social network advertising. The observed improvements highlight that generative models produce valuable synthetic examples that enhance classifiers.

\subsection{Implications for Social Network Advertising}
The findings suggest that advertisers and data scientists can use data augmentation to improve the predictive accuracy of models, potentially leading to more successful targeting and engagement strategies in social media advertising campaigns.

\subsection{Summary of Key Findings}
In summary, this study revealed that data augmentation notably enhances the predictive performance of diverse classifiers. Among the generative models, VAE-generated data typically yields the most substantial performance improvements, followed by GAN, while GMM exhibits a comparatively weaker effect. While augmented data facilitates enhanced model performance, the extent of improvement varies across different models, underscoring the importance of tailored augmentation strategies for each model. These insights add to the body of literature by empirically demonstrating the efficacy of synthetic data augmentation and conducting a comparative examination of various generative models within the domain of social network advertising.

\section{Conclusion}

This study aims to explore the potential of data augmentation in improving predictive modeling of social network advertising data. Through careful methodology and analysis, the research reveals important findings that enrich the body of knowledge about the application of machine learning in digital marketing. Our research shows that data augmentation, specifically through variational autoencoders (VAEs) and generative adversarial networks (GANs), significantly improves the performance of various machine learning classifiers. VAE is considered the most effective augmentation technique, followed by GAN, and Gaussian Mixture Model (GMM) at the bottom. These improvements are measured through standard performance metrics, specifically significant improvements in accuracy.

In summary, this study highlights the transformative potential of data augmentation in the field of social network advertising. It highlights the potential of VAEs and GANs as tools to improve model accuracy and opens endless possibilities for advanced, ethical predictive modeling in digital marketing. As we grapple with the complexities of data-driven advertising, insights from this study will guide future machine learning efforts toward more efficient use of predictive models to understand and engage with social media audiences.

\section{Discussion}
This study warrants a comprehensive discussion that closely links the observed empirical findings to the theoretical framework and research questions posed in the previous article. This discussion contextualizes the impact of data augmentation on the modeling of social network advertising.

\subsection{Impact of Data Augmentation on Classifier Performance}
The positive impact of data augmentation on model performance reaffirms the premise that augmenting training data can improve data scarcity and class imbalance problems. It is worth noting that the variational autoencoder (VAE) appears to be the most effective enhancement method, improving the predictive ability of the classifier, especially for dense neural networks \cite{20,21,22,23}. This superiority may be attributed to VAE's ability to more effectively capture and encode the underlying data distribution, providing a more diverse and representative synthetic dataset. 

\subsection{Theoretical and Practical Implications}
The theoretical implications of these findings provide empirical support for using VAEs and GANs to create synthetic datasets that help build more accurate predictive models \cite{29, 30, 31, 32}. In practice, these findings could influence how data scientists in the field of social network advertising build datasets, balancing the trade-offs between data privacy and the need for large-scale, balanced data.

\subsection{Study Limitations and Future Research Directions}
Limitations of this study must be acknowledged. The scope is limited to a single dataset, which may limit the generalizability of the findings to other types of social network advertising data. Furthermore, although the selected classifiers represent a broad range of machine learning models, they do not cover all algorithms that may be applied in this field.

Future research could verify the generalizability of the findings by applying the same approach to multiple datasets from different social media platforms. There is room for further exploration of more advanced data augmentation techniques and their impact on fairness and bias in model predictions. An in-depth study of the interplay between different types of classifiers and augmentation techniques may provide deeper insights into the best strategies for model training in data-constrained environments.\\

In summary, this paper comprehensively demonstrates the complex interactions between data augmentation techniques and model performance, recognizing the potential of synthetic data to significantly improve predictive modeling of social network advertising.

\renewcommand{\bibfont}{\footnotesize}

\footnotesize{
\bibliographystyle{IEEEtran}
\bibliography{main}
}

\end{document}